# Screening promising CsV$_3$Sb$_5$-like kagome materials from systematic first-principles evaluation

Yutao Jiang[1,2], Ze Yu[1,2], Yuxin Wang[1,2], Tenglong Lu[1,2],

Sheng Meng[1,2,3] *, Kun Jiang[1,3] *, Miao Liu[1,3,4] *

[1]*Beijing National Laboratory for Condensed Matter Physics, Institute of Physics, Chinese Academy of Sciences, Beijing, 100190, China*

[2]*School of Physical Sciences, University of Chinese Academy of Sciences, Beijing 100190, China*

[3]*Songshan Lake Materials Laboratory, Dongguan, Guangdong, 523808, China*

[4]*Center of Materials Science and Optoelectronics Engineering, University of Chinese Academy of Sciences, Beijing, 100049, P. R. China*

*Corresponding author: smeng@iphy.ac.cn, jiangkun@iphy.ac.cn, mliu@iphy.ac.cn

Y. J. and Z. Y. contributed equally to this work.

**Abstract**

CsV$_3$Sb$_5$ kagome lattice holds the promise for manifesting electron correlation, topology and superconducting. However, by far only three CsV$_3$Sb$_5$-like kagome materials have been experimentally spotted. In this work, we enlarge this family of materials to 1386 compounds via element species substitution, and the further screening process suggests that 28 promising candidates have superior thermodynamic stability, hence they are highly likely to be synthesized. Moreover, these compounds possess several identical electronic structures, and can be categorized into five non-magnetic and three magnetic groups accordingly. It is our hope that this work can greatly expand the viable phase space of the CsV$_3$Sb$_5$-like materials for investigating or tuning the novel quantum phenomena in kagome lattice.

**Introduction**

In condensed matter physics, kagome lattice serves as an important quantum material system to manifest/investigate the interplay between electron correlation, topology, and lattice geometry. [1,2] For example, kagome lattice is featuring novel electronic structures, such as the flat band, Dirac cone, Van-Hove points, etc., owing to its special lattice geometry. [3] By far, several materials have been discovered to host the kagome lattices and novel kagome physics, e.g., the Fe kagome lattice within the $Fe_3Sn_2$ shows giant spin-orbit tunability [4,5] and the long sorted magnetic Weyl semimetals were identified in kagome $Co_3Sn_2S_2$ [6–8]. Recently, the newly discovered quasi-2-dimensional kagome materials, which are $KV_3Sb_5$, $RbV_3Sb_5$, and $CsV_3Sb_5$, open a new avenue for connecting the superconductivity with kagome physics [3,9–16], escalating the curiosity of physics community to uncover even more unconventional quantum phenomena from the kagome lattice.

Unfortunately, finding the desired kagome materials is a challenging task. For example, only three $CsV_3Sb_5$-like superconductive kagome materials (as mentioned above) have been experimentally synthesized at the time of writing, limiting the investigation of kagome superconductivity to a fairly small compound space. Hence, the development of quantum physics hinges pretty much on finding a viable material system with low dimensionality as well as various symmetry-breaking instabilities (such as superconductivity).

In order to search for promising kagome materials, this work systematically evaluates 1386 structures of kagome materials with the stoichiometry of $AM_3X_5$ (A for alkali elements, M for transition metals, and X for anionic elements from III A, IV A, V A), which is derived from the $CsV_3Sb_5$ (space group *P6/mmm*). Employing the high-throughput calculations at the density functional theory level as well as the atomly.net [17] materials database, the optimized structure, thermodynamic stability (energy above hull < 5meV/atom), and electronic structures are calculated, and, in the end, 28 promising candidates are popped out from the screening process, including the $CsV_3Sb_5$, $RbV_3Sb_5$, and $KV_3Sb_5$ – the three known compounds that have been experimentally made according to existing literature, validating the reliability of our calculations. Among the 25 new compounds, it is found that the non-magnetic can be grouped into five non-magnetic categories based on their electronic structures. There are also a handful of magnetic kagome materials that fall out of the screening, and they are likely the feasible compounds to materialize the spin-related quantum phenomena in kagome lattice. This work charts a synthesizability "treasure hunt map", from the theory- and data-driven approach, for quantum material community to discover the viable kagome materials.

**Methodology**

The theoretical calculations are performed at Density Functional Theory (DFT) [18] level in a high-throughput mode to evaluate the properties of $AM_3X_5$ kagome system. The DFT calculations are carried out using the Vienna ab initio software package (VASP) [19,20] with the projector augmented-wave (PAW) [21,22] method to describe the nucleus-electron interactions and the generalized gradient approximation (GGA) within the Perdew-Burke-Ernzerhof (PBE) [23] framework to describe the exchange-correlation in-between electrons. The plane-wave cutoff energy is 520 eV, the energy convergence criterion is $5*10^{-6}$ eV, and the K-points grid density of 100 k-points/Å$^{-1}$ is adopted for all the calculations. All the calculations are conducted in a high-throughput mode by employing the standardized parameters to ensure that the calculation results (especially the formation energy) are mutually comparable to each other.

**Results**

In order to search for $CsV_3Sb_5$-like compounds that are highly synthesizable, 1386 structures are generated by employing the $CsV_3Sb_5$ (space group *P6/mmm*) as the structural template, then an element substitution process (as shown in Fig. 1a) is performed by replacing Cs site with (H, Li, Na, K, Rb, Cs), substituting V site with (Ti, Zr, Hf, V, Nb, Ta, Cr, Mo, W, Mn, Tc, Re, Fe, Ru, Os, Co, Rh, Ir, Ni, Pd, Pt), and replacing Sb site with (Si, Ge, Sn, Pb, P, As, Sb, Bi, S, Se, Te), totaling 1386 structures (including the $CsV_3Sb_5$). The purpose of the procedure is to create a large pool of candidate materials for future calculation and screening.

Then the high-throughput calculations are performed with an in-house workflow for structure optimization, high precision static calculation and electronic structure (density of states and energy bands) calculation. The structure optimization process relaxes the cell volume and the position of each atoms in the structure to yield the equilibrium ground state of the given structure at 0K, then the total energy obtained from the static calculation is used to justify the thermodynamic stability of the compound. In practice, the compound is put together with all available phases from the same chemical system to evaluate the phase competition. In order to get as many compounds as available from the same chemical system, we first query all the compounds from the atomly.net [17], which is an open-access materials database, to get all the relevant data. Then the data is plugged into the PhaseDiagram module [24] from the pymatgen [25] to generate the phase diagrams for each chemical system to obtain the energy above hull ($E_{hull}$) of each $AM_3X_5$ kagome material. Basically, the formation energy means the energy released or consumed when one unit of a substance is created under standard conditions from its pure elements. [24,26] The compounds with smaller formation energy (negative value for energy release) are thermodynamically more stable. The most stable compounds open a formation energy convex hull to describe the lower boundary of the formation energies, and $E_{hull}$ simply denotes the distance between formation energy of a compound to the energy convex hull. In short, $E_{hull}$ of a compound quantitatively measures the driving force (in units of energy) to decompose the compounds to the most stable phases. The comprehensive concept of $E_{hull}$ can be found in ref. [27–29] and supplementary materials (Fig.S1). In general, smaller $E_{hull}$ represents better stability. Normally, nonmetallic compounds with $E_{hull}$ < 50 meV/atom or alloys/metals with $E_{hull}$ < 10 meV/atom are likely to exist in real world.

Fig. 1 (d-i) show the $E_{hull}$ of $AM_3X_5$ compounds from our calculations. It can be observed that $HM_3X_5$, $LiM_3X_5$ compounds are all thermodynamically unstable as their $E_{hull}$ are all larger than 60 meV/atom. In the rest chemical space, $E_{hull}$ of 28 compounds are under 5meV/atom (Table 1), $E_{hull}$ of 38 compounds are less than 20meV/atom (see Table S1 for complete list), and $E_{hull}$ of 90 compounds are smaller than 50meV/atom (see Table S2 for complete list).

Table 1 presents the full list of the 28 compounds that fall out of the $E_{hull}$<5meV/atom screening. It is found that three known compounds, which are $CsV_3Sb_5$, $RbV_3Sb_5$, and $KV_3Sb_5$, show up on the list, suggesting that the screening treatment is fairly effective and accurate. It can be observed that, the stable $AM_3X_5$ structures can be found in $AV_3Sb_5$ (A=Na, K, Rb, Cs), $ANb_3Bi_5$ (A=K, Rb, Cs), $ATi_3Bi_5$ (A=Na, K, Rb, Cs), $AMn_3Sb_5$ (A=K, Rb, Cs), $AHf_3Bi_5$ (A=K, Rb, Cs), $CsTi_3Te_5$, $CsTi_3Sb_5$, $CsV_3Bi_5$, $CsFe_3Ge_5$, $CsFe_3Sn_5$, $CsZr_3Bi_5$, $CsZr_3Te_5$, $CsHf_3Te_5$, $CsTa_3Sb_5$, $CsRh_3Pb_5$, and $CsPd_3Pb_5$. It is also found that the stability of

$AM_3X_5$ decreases sequentially when A site is occupied by H, Li, Na, K, Rb, Cs, and the later transition metal on M site is unlikely to yield stable $AM_3X_5$. Cs on A site can stabilize the structure as 16 out of 28 stable compounds in Table 1 are Cs-containing compounds, and it seems that the stability of $AM_3X_5$ is largely determined by the M-X layer as V-Sb, Nb-Bi, Ti-Bi, Mn-Sb, and Hf-Bi combinations are more stable. Overall, the viable $AM_3X_5$ kagome lattices are limited to a small phase space based on our screening, and Figure 1 and Table 1 chart a likelihood 'map' for discovering the viable $CsV_3Sb_5$-like compounds.

Since the electronic structures, especially the flat-band (FB), Dirac-point (DP), and Van Hove singularity (VHS), are fairly important properties of kagome lattice, the electronic structures of all the $CsV_3Sb_5$-like compounds (1386 of them in total) are calculated at GGA-PBE level without considering the spin orbital coupling. It is found that the $AM_3X_5$ can be classified into five non-magnetic groups and three magnetic groups according to their energy bands and Fermi surface. Figure 2 shows the representative compounds of the five non-magnetic types, and their electronic structures. The $CsV_3Sb_5$ is the most known type, and the DP is below the Fermi surface and VHS is close to or at the Fermi surface. Other than $AV_3Sb_5$ (A = Na, K, Rb, Cs), $CsV_3Bi_5$ is also in this group. For the $RbNb_3Bi_5$-type, the DP is below the Fermi surface, the VHS is close to or at the Fermi surface, but the bands, except the VHS band, at the M point open up a gap. $KNb_3Bi_5$, $RbNb_3Bi_5$, $CsNb_3Bi_5$, and $CsTa_3Sb_5$ belong to this type of compound. The $CsTi_3Bi_5$-type shots up both the DP and VHS as those compounds have less valence electrons than that of $CsV_3Sb_5$-type, hence it can be regarded as the hole-doped $CsV_3Sb_5$-type and the Fermi surface is moved downward to some extent. $ATi_3Bi_5$ (A=Na, K, Rb, Cs), $AHf_3Bi_5$ (A=K, Rb, Cs), $CsTi_3Sb_5$, and $CsZr_3Bi_5$ can be classified into this electronic structure type. In $CsTi_3Te_5$-type compounds, the VHS band is vanished and the DP is located fairly close to the Fermi surface, forming a simple Fermi surface in reciprocal space as there are fewer bands around. $CsTi_3Te_5$, $CsZr_3Te_5$, $CsHf_3Te_5$, are the similar type of compounds. $CsRh_3Pb_5$-type compounds are more three-dimensional losing the typical feature of two-dimensional Kagome lattice. $CsRh_3Pb_5$ still has the VHS band near the Fermi surface but lacks the DP bands, and $CsRh_3Pb_5$, $CsPd_3Pb_5$ are this type of compounds.

For the Mn- and Fe-containing $AM_3X_5$ structures, the compounds are spin-polarized and show ferromagnetism as shown in Fig. 3. There are only a few of this type of compounds, those are $AMn_3Sb_5$ (A=K, Rb, Cs), $CsFe_3X_5$ (X=Ge, Sn). It is fairly special that the kagome sites in those compounds are spin-polarized, making them the viable material systems to materialize the spin-related phenomenon. [1,2,30,31] At the collinear level of treatment for the system, it is observed that the Mn or Fe kagome sites hold the prominent amount of magnetic moment, e.g. the magnetic moment of Mn sites in $CsMn_3Sb_5$ is around 2.5-2.7 $\mu_B$/atom while the total magnetic moment of the compounds is around 7.5-8.0 $\mu_B$ per formula.

Because of the non-degenerated orbitals caused by the spin-polarization, the Fermi surface in the reciprocal space manifests much more complicated structures, and the DP and VHS are also split into two sets. Hence, our work hints that it is likely to have spin-polarized $CsV_3Sb_5$-like kagome materials, and a drilled-down study is worth conducting.

**Discussions**

Based on our screening, we are offering the community a ranked list for discovering new compounds that share the same crystal structure with $CsV_3Sb_5$. We indeed obtain a compound list of $CsV_3Sb_5$-like compounds, and provide a systematic evaluation of formation energy landscape for them. Considering that the formation energy of alloy is generally small compared to the ionic and covalent compounds, and the formation energy between the compounds is still relatively close to each other, therefore, it is generally difficult to synthesize the intermetallic phases even when they are predicted to be stable.

The data that we utilized to generate the phase diagram cover a very large compound space as most ICSD [32] and materials project [33] structures are included [Here, you may need add ICSD and material project ref. or website]. More than that, we have generated and calculated ~100k new structures by replacing the element species. The large compound space can ensure the accurate gauging of the $E_{hull}$ of compounds, but there is still a small chance to overlook the most stable compounds that can stretch the hull to even lower energies. In such case, the $E_{hull}$ of most compounds can be lifted up a little, hence the value of the $E_{hull}$ may increase a little when the database grows larger.

The Table 1, S1 and S2 serve as synthesis guides for exploring the $CsV_3Sb_5$-like compounds, if the pure phases of those predicted compounds are hard to make, it is highly possible that the doped compounds can be experimentally realized employing the knowledge from Fig. 1. For example, hole- or electron-doping to $CsV_3Sb_5$ compounds could probably be experimentally realized by replacing a small amount of the V with Ti or Ni [34,35], and substituting small amount of the V with Mn or Fe can possibly introduce magnetism into these material systems. Fig. 2 and 3 provide guides on how the electronic structures can be tuned when mixing the metal elements. Although, the main text of the paper shows the 28 most stable compounds, those compounds in Table S1 also have good-to-moderate stability, and are worth looking at. For example, the $CsV_6Sb_6$, which has a $E_{hull}$ of ~12 meV/atom based on our calculation, has been experimentally made [36].

The electronic structures of all the 28 compounds in Table 1 can be found in the supplementary material (Fig. S2). The calculated data for all the 1386 compounds is open-accessible from atomly.net [17] (for people who has a user account).

## Conclusions

In summary, this work widens the knowledge of $CsV_3Sb_5$-family through high-throughput calculation and systematic evaluation. Via the element replacement, a large chemical space of the $CsV_3Sb_5$-like materials (1386 in total) are calculated at the GGA-PBE level employing the DFT, the systematic evaluation tells us that there are at least 28 compounds having excellent thermodynamic stability, and therefore it looks forward to a further experimental confirmation. The electronic structures of those materials add possibilities for rational band structure engineering of kagome quantum materials. It is our hope that useful new kagome materials can be experimentally discovered for quantum physics community based on our prediction.

## Acknowledgements

We would acknowledge the financial support from the Chinese Academy of Sciences CAS (No. ZDBS-LY-SLH007, and No. XDB33020000) and the National Natural Science Foundation of China (Grant No. 12174428). The computational resource is provided by the Platform for Data-Driven Computational Materials Discovery of the Songshan Lake Laboratory. We also thank Youguo Shi for discussion.

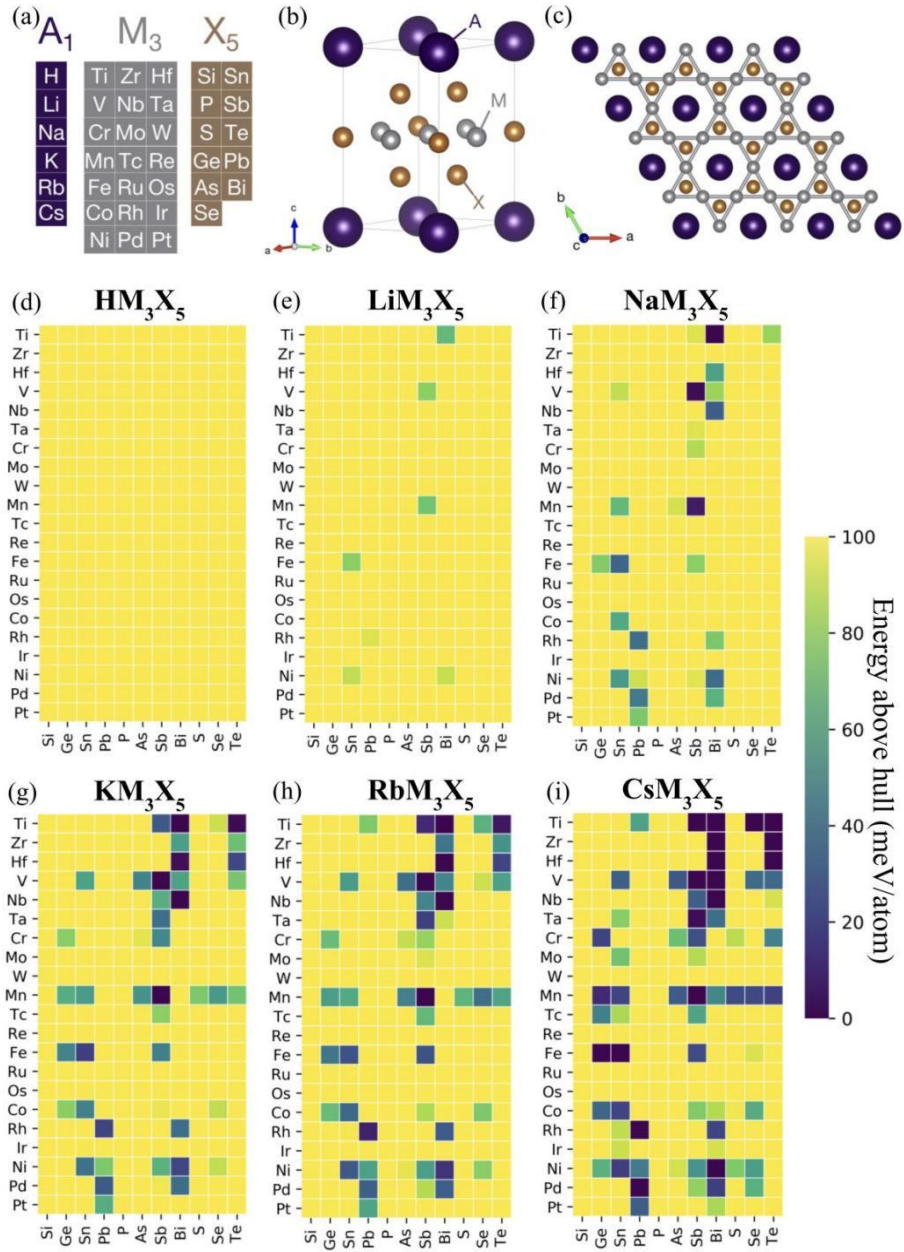

Figure 1. The structure and thermodynamic stability of the kagome compounds. (a-c) the structure and the schematic of the element substitution strategy for generating $CsV_3Sb_5$-like compounds, and the transition metal forms the kagome lattice in structure. (d-i) The thermodynamic stability (represented quantitatively as $E_{hull}$) for 1386 compounds in a heat map plotting mode. Each box shows a $AM_3X_5$ compound, and the darker color the more stable the compound is. Compounds with $E_{hull}$ greater than 100 meV/atom are plotted in yellow.

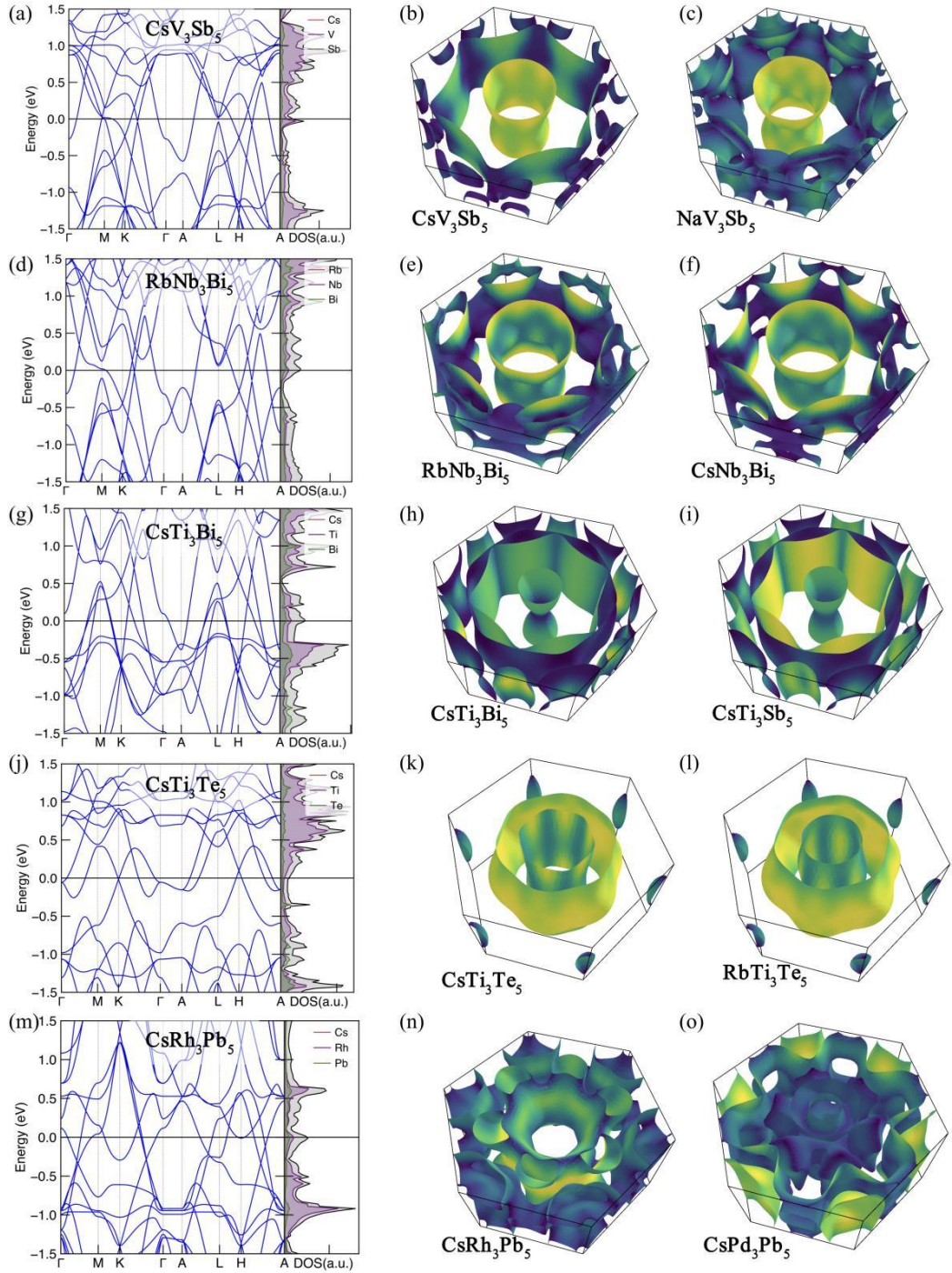

Figure 2. Electronic structures of selected non-magnetic compounds with $E_{hull}$<5 meV/atom. The compounds are categorized into five dedicated kinds, as (a)-(c), (d)-(f), (g)-(i), (j)-(l), (m)-(o), according to their band structures, density of states and Fermi surfaces. Compounds in the same group have very similar electronic structures.

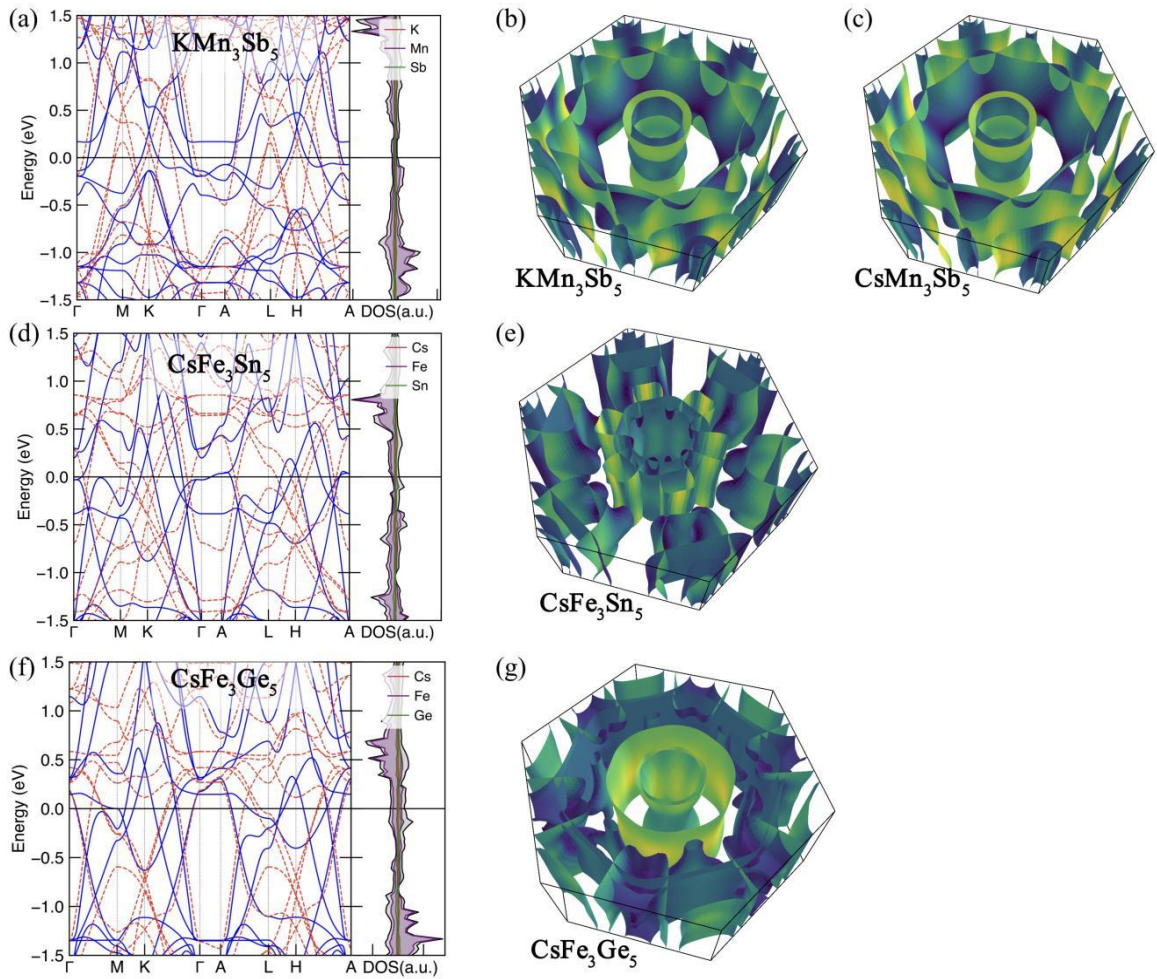

Figure 3. Electronic structures of magnetic compounds with $E_{hull}$<5meV/atom. (a)-(c), (d) and (e), (f) and (g) are band structures, density of states and Fermi surfaces of representative compounds.

Table 1. List of 28 compounds with good thermodynamic stability ($E_{hull}$<5 meV/atom). The compound formula, ID in the database, Ehull value and the lattice parameters are provided.

| Formula | ID (atomly.net) | $E_{hull}$ (meV/atom) | a(Å) | c(Å) | Magnetization ($\mu_B$/formula) |
|---|---|---|---|---|---|
| NaTi$_3$Bi$_5$ | 1000299281 | 0 | 5.78 | 9.18 | |
| NaV$_3$Sb$_5$ | 1000299291 | 2 | 5.46 | 8.91 | |
| KTi$_3$Bi$_5$ | 1000299512 | 0 | 5.79 | 9.72 | |
| KV$_3$Sb$_5$ | 1000299522 | 0 | 5.48 | 9.37 | |
| KMn$_3$Sb$_5$ | 1000299544 | 0 | 5.43 | 9.26 | FM($\mu$=7.75) |
| KNb$_3$Bi$_5$ | 1000299600 | 0 | 5.89 | 9.59 | |
| KHf$_3$Bi$_5$ | 1000299666 | 3 | 6.08 | 9.45 | |
| RbTi$_3$Bi$_5$ | 1000299743 | 0 | 5.79 | 9.89 | |
| RbV$_3$Sb$_5$ | 1000299753 | 0 | 5.50 | 9.38 | |
| RbMn$_3$Sb$_5$ | 1000299775 | 0 | 5.44 | 9.44 | FM($\mu$=7.73) |
| RbNb$_3$Bi$_5$ | 1000299831 | 0 | 5.89 | 9.71 | |
| RbHf$_3$Bi$_5$ | 1000299897 | 0 | 6.08 | 9.59 | |
| CsTi$_3$Sb$_5$ | 1000299973 | 0 | 5.68 | 9.81 | |
| CsTi$_3$Bi$_5$ | 1000299974 | 0 | 5.83 | 9.92 | |
| CsTi$_3$Te$_5$ | 1000299981 | 0 | 6.10 | 8.91 | |
| CsV$_3$Sb$_5$ | 1000299984 | 0 | 5.49 | 9.89 | |
| CsV$_3$Bi$_5$ | 1000299985 | 0 | 5.64 | 10.15 | |
| CsMn$_3$Sb$_5$ | 1000300006 | 0 | 5.44 | 9.69 | FM($\mu$=7.70) |
| CsFe$_3$Ge$_5$ | 1000300020 | 0 | 4.89 | 9.73 | FM($\mu$=4.99) |
| CsFe$_3$Sn$_5$ | 1000300021 | 0 | 5.26 | 10.53 | FM($\mu$=6.50) |
| CsZr$_3$Bi$_5$ | 1000300051 | 0 | 6.11 | 10.09 | |
| CsZr$_3$Te$_5$ | 1000300058 | 0 | 6.41 | 9.07 | |
| CsNb$_3$Bi$_5$ | 1000300062 | 0 | 5.91 | 9.87 | |
| CsRh$_3$Pb$_5$ | 1000300110 | 0 | 5.71 | 9.74 | |
| CsPd$_3$Pb$_5$ | 1000300121 | 0 | 5.81 | 9.61 | |
| CsHf$_3$Bi$_5$ | 1000300128 | 0 | 6.09 | 10.05 | |
| CsHf$_3$Te$_5$ | 1000300135 | 0 | 6.35 | 8.76 | |
| CsTa$_3$Sb$_5$ | 1000300138 | 2 | 5.81 | 9.60 | |